**Josephson junction π-0 transition induced by orbital hybridization in a double quantum dot**

*Rousan Debbarma,[1,*,†] Athanasios Tsintzis,[1] Markus Aspegren,[1] Rubén Seoane Souto,[3,4]*

*Sebastian Lehmann,[1] Kimberly Dick,[1,2] Martin Leijnse,[1] Claes Thelander[1,‡]*

[1]Division of Solid State Physics and NanoLund and [2]Center for Analysis and Synthesis, Lund University, S-221 00 Lund, Sweden

[3]Departamento de Física Teórica de la Materia Condensada, Condensed Matter Physics Center (IFIMAC) and Instituto Nicolás Cabrera, Universidad Autónoma de Madrid, 28049 Madrid, Spain

[4]Instituto de Ciencia de Materiales de Madrid (ICMM), Consejo Superior de Investigaciones Científicas (CSIC), Sor Juana Inés de la Cruz 3, 28049 Madrid, Spain.

[*]Present address: Department of Chemical Engineering, Indian Institute of Technology Delhi, Hauz Khas, New Delhi, Delhi 110016, India

*Abstract*

In this work, we manipulate the phase shift of a Josephson junction using a parallel double quantum dot (QD). By employing a superconducting quantum interference device, we determine how orbital hybridization and detuning affect the current-phase relation in the Coulomb blockade regime. For weak hybridization between the QDs, we find π junction characteristics if at least one QD has an unpaired electron. Notably the critical current is higher when both QDs have an odd electron occupation. By increasing the inter-QD hybridization the critical current is reduced, until eventually a π-0 transition occurs. A similar transition appears when detuning the QD levels at finite hybridization. Based on a zero-bandwidth model, we argue that both cases of phase-shift transitions can be understood considering an increased weight of states with a double occupancy in the ground state and with the Cooper pair transport dominated by local Andreev reflection.

*Introduction*

Josephson junctions (JJs) with embedded quantum dots (QDs) are ideal systems for exploring the interplay between electron correlation and Josephson effects [1–4]. Recent research has focused on controlling symmetry-breaking mechanisms in Cooper pair transport, as they can give rise to dissipation-free diode behavior and a fixed-phase offset in the ground state [5–7]. Insights into different Cooper pair transport processes in a JJ is typically gained by studying its current-phase relationship (CPR). The Josephson current, $I_J$, for low-transmission junctions, relates to the critical current $I_C$, and phase $\phi$, across the JJ as $I_J = I_C \sin(\phi + \alpha)$, where $\alpha$ is a phase shift that depends on the junction properties. In a model proposed by Kulik, tunneling processes that preserve and reverse the spin-ordering of the Cooper-pairs are both present in a JJ, where the CPR is determined by their net effect [8]. When spin-order preserving processes dominate, $\alpha = 0$ (0 junction), otherwise $\alpha = \pi$ ($\pi$ junction) and $I_J = -I_C \sin(\phi)$ [8–10]. 0-$\pi$ transitions in a JJ can be caused by various factors including Coulomb interactions, subgap states, orbital parities, quasiparticle densities and the Kondo effect [1,11–14].

In systems with double QDs (DQDs), it is possible to control the strengths of orbital and spin interactions over a wide range, which is taken advantage of in spin-qubit implementations [15–21]. DQDs coupled to a joint superconducting contact offer the possibility to generate spin-correlated electron pairs [22–25]. Embedded inside a JJ, DQDs have also been used to explore how $I_J$ depends on spin and orbital states [26–28]. Here, the CPR can be determined by using a superconducting quantum interference device (SQUID) [2,29]. However, for parallel DQDs CPRs have been primarily obtained in theoretical studies, whereas most experimental studies instead indirectly determine 0-$\pi$ transitions based on changes in critical currents [13,27,30,31]

In this Letter, we test the Kulik picture in a DQD system where the relative contribution of processes that preserve or reverse the spin order can be tuned using electrostatic gates. By embedding a parallel DQD in a SQUID, we are able to study the effect of interdot hybridization and level detuning on the phase shifts of $I_J$. By increasing the interdot hybridization, going from weak to strong interdot tunnel coupling, we find a $\pi$-0 transition for electron configurations where each QD provides an unpaired spin. For an intermediate coupling, the phase shift within such a charge configuration is also found to change with level detuning ($\delta$). Our experimental results are complemented with zero bandwidth (ZBW) calculations [13,32] showing that $\pi$-0 transitions can be attributed to the suppression of crossed Andreev reflection (CAR) and correlated to an increasing relative weight of QD double occupancy in the ground state due to interdot hybridization and detuning.

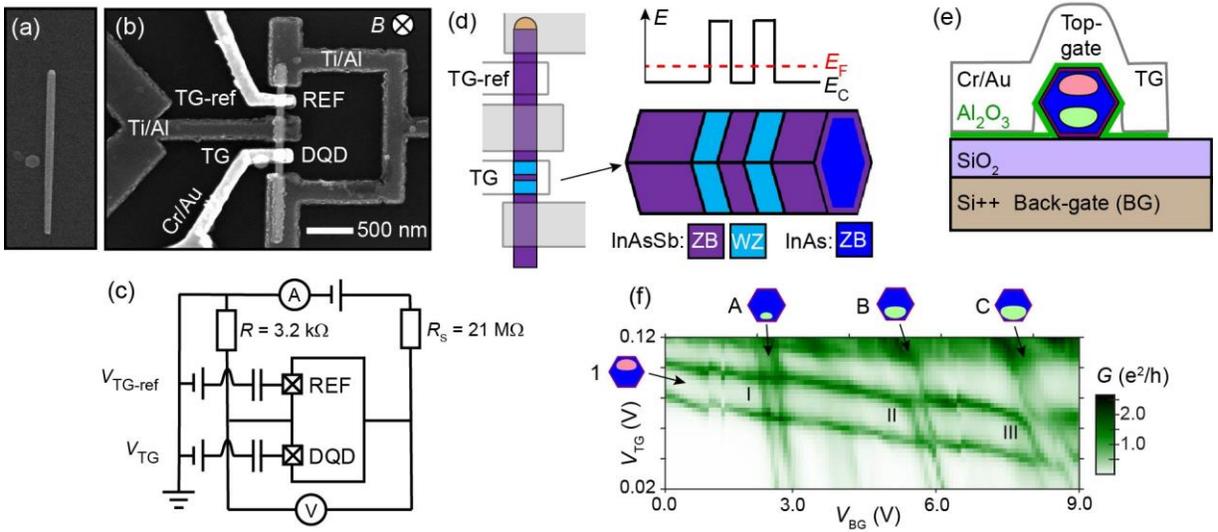

Figure 1: (a) SEM image of the nanowire. (b) SQUID with two JJs, where one contains a nanowire section with two wurtzite (WZ) barrriers (labelled DQD) and the other a zinc blende (ZB) segment (labelled REF) of the same nanowire. The magnetic field, $B$, is applied perpendicular to the plane of the device for CPR measurements. (c) Schematic circuit diagram of the four-probe setup used for supercurrent measurements. (d) Schematics of the nanowire device and the conduction band energies of the crystal phases. (e) Cross-section schematic of the nanowire device showing the top-gate and back-gate controlled DQD. (f) Charge stability diagram as a function of the DQD top-gate ($V_{TG}$) and back gate ($V_{BG}$) for a source-drain voltage ($V_{SD}$) of 0.43 mV, showing three different DQD crossings. Orbital crossings I, II, III occur when an orbital with strong electrostatic coupling to the top-gate interacts with orbitals with strong coupling to the back gate. The reference channel is in the off state ($V_{TG\text{-ref}} < 0$).

*SQUID fabrication and measurement setup*: In this work, we use crystal-phase-defined nanowire QDs where the electrons are confined in a very short zinc blende (ZB) segment surrounded by wurtzite (WZ) barriers in an InAs-InAsSb core-shell configuration [21,28]. The ZB QD segment is 5-10 nm long surrounded by WZ barriers that are 20 nm thick with proximitized ZB segments 150-180 nm long on each end. First, gate electrodes are formed by electron-beam lithography, followed by atomic-layer deposition of 3 nm $Al_2O_3$ and evaporation of Cr/Au (5/95 nm). Next, Ti/Al (5/90 nm) electrodes are processed, including a superconducting loop. Figure 1b shows an SEM image of the SQUID with two JJs, where one hosts the DQD and the other a reference ZB nanowire segment.

All measurements were performed in a dilution refrigerator at a base temperature of 15 mK. For the supercurrent measurements, we employ a four-probe setup as shown in the circuit diagram in

Fig 1c. The electron population in both arms of the SQUID are controlled with gate electrodes and can be switched on (conducting) and off (non-conducting). A large series resistance ($R_S = 21$ M$\Omega$) is used for current-biased supercurrent measurements. To obtain CPR, both arms of the SQUID are in an on state such that the two JJs interfere. All the SQUID measurements were done in the Coulomb blockade regime, thus involving Cooper-pair transport mechanisms of fourth-order or higher in the lead-QD tunnel coupling [2,10]. In case of voltage-bias measurements, such as charge-stability diagrams, we use a two-probe setup where the reference arm of the SQUID is in an off state ($V_{\text{TG-ref}} < 0$) and without $R_S$.

*Results*: At sufficiently low electron occupancy, the crystal-phase defined QD breaks into a parallel DQD configuration [19] where the orbital energies of the two QDs are controlled using the top gate (TG) and the back gate (BG) (Fig. 1e). The charge stability diagram in Fig. 1f shows the formation of a parallel DQD where orbital-1, strongly coupled to TG, interacts with three orbitals A, B, and C, strongly coupled to the BG leading to the formation of crossings I, II and III with increasing interdot tunnel coupling.

The SQUID measurement in Fig. 2a shows the differential resistance ($dV/dI$) of the device as a function of magnetic field ($B$) and bias current ($I$). The switching current ($I_{\text{SW}}$) at a given $B$-field is extracted from the boundaries of the zero-resistance region and fitted (red curves) to a sinusoidal SQUID equation [33]. From here on, all the CPR plots will show extracted $I_{\text{SW}}$ and corresponding SQUID equation fit. Figure 2b shows the CPR for a π junction (red) and 0 junction (blue). To rule out spurious effects such as magnetic hysteresis, we carried out two measurements per $B$-field, alternating the gate-voltage between a spin-1/2 and a spin-0 state in orbital 1. We first present the experimental results for the three orbital crossings followed by theoretical results and discussion.

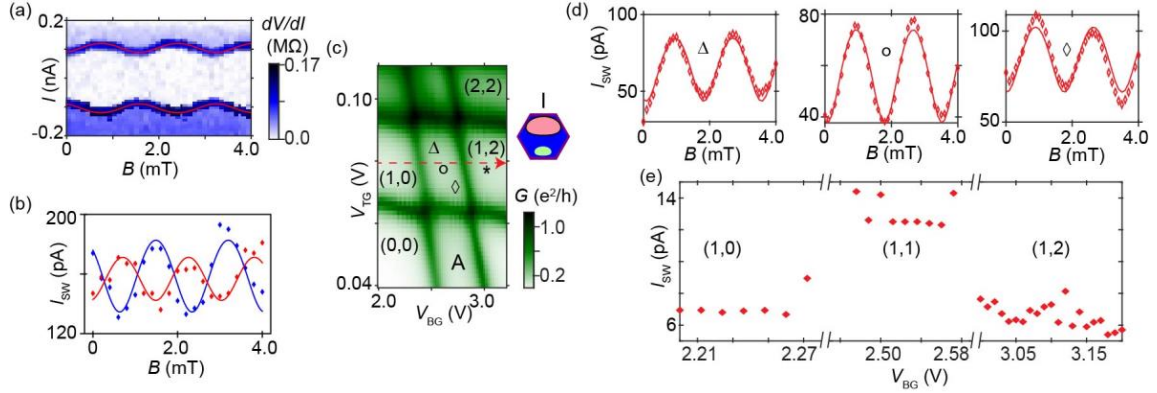

Figure 2: (a) Differential resistance ($dV/dI$) across the SQUID for a spin-1/2 state (* in panel c) as a function of bias current and $B$-field. The switching current ($I_{SW}$) is given by the boundary of the white zero-resistance supercurrent region. The red curve is a sinusoidal fit for the extracted $I_{SW}$. (b) $I_{SW}$ and the SQUID equation fit for spin-0 (blue) and spin-1/2 (red) states in orbital-1 measured by alternating $V_{TG}$ for each $B$. The spin states are away from the DQD crossings ($V_{BG}=0$ V in Fig. 1(f)). (c) Charge stability diagram for the weakly coupled orbital crossing I at $V_{SD}= 0.23$ mV. The numbers in parentheses indicate the charge states of the DQD. (d) CPR for various points inside the (1,1) honeycomb (positions indicated by the symbols in panel c). (e) Extracted $I_{SW}$ along the red vector in panel c with the reference channel in off-state ($V_{TG\text{-ref}} < 0$).

*Weak interdot coupling*: We start by discussing the results for orbital crossing I, which we refer to as the weak interdot coupling case. The charge stability diagram for orbital crossing I in Fig. 2c shows no avoided crossings and thus represents a DQD with weak interdot tunnel coupling [34]. Away from the orbital crossing, we find no phase shift in the CPR for the spin-0 states, (0,0) and (0,2). However, for the spin-1/2 states, (1,0) and (0,1), the CPR shows a π shift. Corresponding π shifts for single isolated orbitals have been shown in previous studies [2,35]. Here, we find that a π shift can be obtained in a DQD even though one QD has even spin pairing, such as (1,0), (0,1) and (1,2), as long as the filled (or next empty) orbital is energetically far away. We note that once orbital-1 is filled, the CPRs exhibit no phase shift. A possible explanation is an increase in the coupling between the DQD and the superconducting leads once orbital-1 is fully occupied, which becomes the dominant channel for Cooper pair transport [36,37]. The CPRs for isolated orbitals are presented in the Supplemental Material [38].

Inside the (1,1) honeycomb, where each QD hosts an unpaired electron, the phase shift is independent of level detuning, as shown in the CPRs with π shifts in Fig. 2d. We understand this by considering the two weakly coupled QDs as two independent spin-1/2 channels carrying negative (π-shifted) $I_J$. The claim of independent channels is supported by the $I_{SW}$ dependence on the charge states as shown in Fig. 2e, where orbital-1 energy level is kept fixed in the Coulomb blockade regime while varying the energy level of orbital-A. We find that $I_{SW}$ inside the (1,1) honeycomb with two unpaired electrons is higher than in the (1,0) and (1,2) configurations with only one unpaired electron.

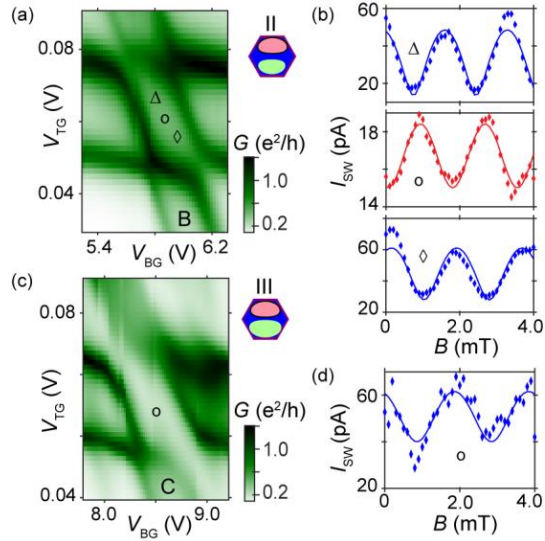

Figure 3: (a) Charge stability diagram for orbital crossing II with intermediate interdot tunnel coupling at $V_{SD}$ = 0.23 mV. (b) CPR for various points inside the (1,1) honeycomb showing the dependence of the phase-shift on level detuning. (c) Charge stability diagram for the strongly coupled orbital crossing III ($V_{SD}$ = 0.23 mV) and (d) CPR at the center of the (1,1) honeycomb with no phase-shift.

*Intermediate and strong interdot coupling*: The charge stability diagram for orbital crossing II in Fig. 3a shows avoided crossings indicating a tunnel coupling between the two QDs, which we will refer to as the intermediate coupling case. We estimate the interdot tunnel coupling, $t \sim 0.8$ meV, based on the charging energy, $E_C$ = 8 meV, of the top QD [38]. In contrast to the weak coupling case, the CPR inside the (1,1) honeycomb for the intermediate case (Fig. 3b) depends on the level detuning ($\delta$) of the two QDs, with a $\pi$ shift at the center ($\delta$ = 0) and no phase shift for $\delta$ above a certain threshold. Another difference is that the amplitude of $I_{SW}$ for the intermediate case at $\delta = 0$ is one order of magnitude smaller than for the weak coupling case (explained later). We note a slight phase shift between the two CPRs at finite detuning (∆ and ◊) which is due to a small hysteretic contribution from the magnet and not due to the emergence of an arbitrary phase shift (such as reported by Szombati et al [7]). This is confirmed by performing measurements such as shown in Fig. 2b.

The charge stability diagram for the case of strong coupling, Fig. 3c, exhibits even larger avoided crossings, where we estimate $t \sim 1.6$ meV (~0.2 $E_C$ of top QD). The CPR inside the (1,1) honeycomb for δ = 0 shows no phase shift unlike the π shifts observed for the weak and intermediate coupling cases. The data here is noisy because of charge fluctuations visible in the charge stability diagram, possibly linked to the higher $V_{BG}$. We note that the CPR for the (0,1) configuration exhibits no phase shift likely because of stronger coupling of the orbital with the SC leads whereas the (1,0) and (1,2) configurations behave as π junctions (see [38]).

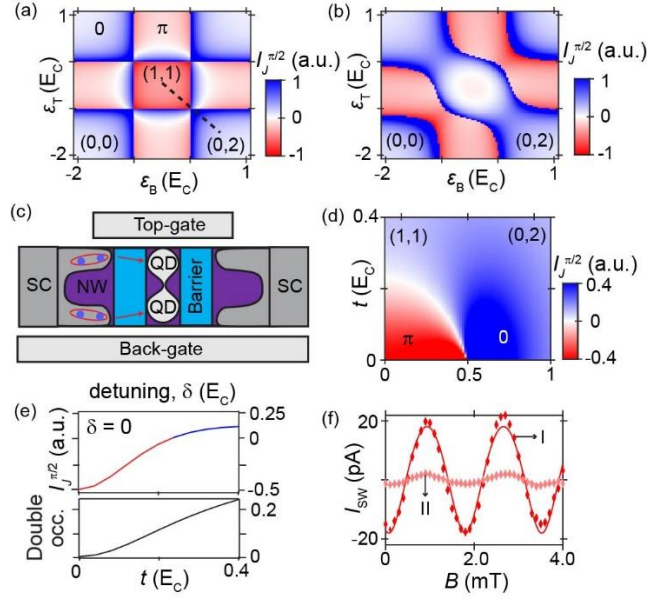

Figure 4: (a) ZBW model calculation of Josephson current $\left(I_J^{\pi/2}\right)$ as a function of QD level energies for $t$ = 0 showing higher negative $I_J^{\pi/2}$ inside the (1,1) honeycomb compared to (1,0) and (1,2). (b) $I_J^{\pi/2}$ plot for $t$ = 0.2 $E_C$ showing a negative value with low magnitude at the center of the honeycomb. (c) Schematics showing two separate modes in the proximitized SC and local tunneling of Cooper pairs. (d) $I_J^{\pi/2}$ as a function of δ and $t$ (dotted line in panel a). (e) $I_J^{\pi/2}$ and double occupancy probability as a function of $t$ at the center of the (1,1) honeycomb for δ = 0. The increasing double occupancy weight leads to a π-0 transition. (f) CPR at the center of the honeycomb for orbital crossings I and II after the subtraction of reference segment $I_{SW}$.

*Theoretical calculation and comparison with experiment*: We calculated the zero-temperature Josephson current ($I_J^{\pi/2}$) through a DQD using a ZBW model [13,32] by taking the derivative of the ground state energy as a function of $\phi$ at $\phi = \pi/2$ (see [38] for details). $I_J^{\pi/2}$ gives an approximation of the critical current and $I_{SW}$. To reach agreement with the experiment, we adapted the model to suppress CAR, such that the QDs couple to different modes in the proximitized leads [38]. The presence of separate modes in the leads has been inferred before in similar DQD structures, as it is a prerequisite for observation of orbital Kondo physics [39]. We propose that the positive voltages applied to the two gates, used to induce the two QDs, also increases the carrier concentration on opposite flats of the nanowire in the leads, resulting in a dominance of local Andreev reflection processes (Fig. 4c). By having separate gates to the leads, or superconductors that extend to the tunnel barriers, we believe it will be possible to tune transport into a regime that favors CAR. Fig. 4a shows $I_J^{\pi/2}$ as a function of the QD levels for $t = 0$. It shows good qualitative agreement with the results for the weak coupling case with negative $I_J^{\pi/2}$ ($\pi$ shift) in the (1,1) honeycomb and a magnitude that is higher than in (1,0) and (1,2) configurations. As mentioned earlier, this can be understood by considering the two QDs as two independent spin-1/2 channels. Fig. 4b shows $I_J^{\pi/2}$ for $t = 0.2\ E_C$, which agrees well with the experimental results for the intermediate coupling case, albeit with an estimated $t \sim 0.1\ E_C$. At the center of the honeycomb where $\delta = 0$, $I_J^{\pi/2}$ is negative with magnitude lower than for the case of $t = 0$.

Fig. 4d shows $I_J^{\pi/2}$ in the two-electron regime as a function of $t$ and $\delta$ as the DQD levels are detuned from the center of the (1,1) honeycomb towards the doubly occupied (0,2) configuration. The $I_J^{\pi/2}$ plot shows a 0-$\pi$ separation, with a gradual transition between the two regions separated by low $I_J^{\pi/2}$ regions. In an intuitive picture, this can be explained by the presence of both positive (0) and negative ($\pi$) Josephson currents [8,9]. At the transition region, the two contributions are similar in magnitude and cancel each other. Such gradual phase and magnitude transitions are only observed for non-zero $t$. The relative weight of states with double occupancy in the ground state increases with both $\delta$ and $t$. For $t > 0$, the (1,1) singlet couples to the (0,2) and (2,0) singlets [30]. As previous studies have shown, doubly occupied spin-0 states behave as 0 junctions whereas singly occupied spin-1/2 states behave as $\pi$ junctions [2,10]. This effect can also be seen in Fig. 4e where we plot $I_J^{\pi/2}$ and the double occupancy probability at the center of the (1,1) honeycomb for $\delta = 0$ as a function of $t$.

The picture proposed by Kulik [8], suggesting a competition between processes that preserve or reverse the spin order, provides a qualitative understanding of our experimental results. The increased weight of double occupancy with increasing interdot coupling affects the phase-shift and magnitude of $I_{SW}$ even at the center of the (1,1) honeycomb. In Fig. 4f, we plot the CPR at the center of the (1,1) honeycomb for crossings I and II after subtracting $I_{SW}$ in the reference arm. The considerably reduced supercurrent in crossing II indeed points to an increasing spin-preserving component. Detuning within the (1,1) honeycomb for crossing II initially leads to a balance between the two components, but eventually favoring a positive $I_J$ (Fig. 3b). As the interdot coupling is further increased in crossing III, the center of the honeycomb transitions to a 0 junction (Fig. 3d). The Kulik picture also explains the observed phase shift and magnitude in the (1,2) regime for the weak interdot coupling case (see Fig S1c).

*Conclusion*: In summary, we have experimentally determined the current-phase relation of a Josephson junction hosting a parallel-coupled DQD, focusing on the effects of interdot tunnel coupling and level detuning. Using a SQUID, we observed a π-0 transition with increasing QD hybridization and probability of double occupancy. From comparison with ZBW calculations, we propose that the Cooper pair transport is dominated by local Andreev reflection through separate modes in the proximitized NW contacts. The data can be understood by considering the transport process as a competition between negative and positive Josephson currents carried by singly occupied spin-1/2 states and doubly occupied spin-0 states, respectively. The ability to control the phase shift via spin states of individual QDs in a DQD JJ opens up the potential for implementing a Josephson phase battery with a controlled and arbitrary fixed-phase offset [40–43].

ACKNOWLEDGMENT

This work was supported by the Knut and Alice Wallenberg Foundation, the Swedish Research Council, the Crafoord Foundation, and NanoLund. The computations were enabled by resources provided by the National Academic Infrastructure for Supercomputing in Sweden (NAISS) at PDC, the Center for High-Performance Computing at the Royal Institute of Technology (KTH). R.S.S. acknowledges funding support from Spanish CM "Talento Program" Project No. 2022-T1/IND-24070 and the European Union's Horizon 2020 research and innovation program under the Marie Sklodowska-Curie grant agreement No. 10103324.

[†] rousan@chemical.iitd.ac.in

[‡] claes.thelander@ftf.lth.se

# References

[1] E. Vecino, A. Martín-Rodero, and A. L. Yeyati, *Josephson Current through a Correlated Quantum Level: Andreev States and π Junction Behavior*, Phys Rev B **68**, 035105 (2003).

[2] J. A. van Dam, Y. V. Nazarov, E. P. A. M. Bakkers, S. De Franceschi, and L. P. Kouwenhoven, *Supercurrent Reversal in Quantum Dots*, Nature **442**, 667 (2006).

[3] H. I. Jørgensen, T. Novotný, K. Grove-Rasmussen, K. Flensberg, and P. E. Lindelof, *Critical Current 0−π Transition in Designed Josephson Quantum Dot Junctions*, Nano Lett **7**, 2441 (2007).

[4] R. Delagrange, R. Weil, A. Kasumov, M. Ferrier, H. Bouchiat, and R. Deblock, *0- π Quantum Transition in a Carbon Nanotube Josephson Junction: Universal Phase Dependence and Orbital Degeneracy*, Phys Rev B **93**, 195437 (2016).

[5] J.-X. Lin, P. Siriviboon, H. D. Scammell, S. Liu, D. Rhodes, K. Watanabe, T. Taniguchi, J. Hone, M. S. Scheurer, and J. I. A. Li, *Zero-Field Superconducting Diode Effect in Small-Twist-Angle Trilayer Graphene*, Nature Physics 2022 18:10 **18**, 1221 (2022).

[6] B. Pal et al., *Josephson Diode Effect from Cooper Pair Momentum in a Topological Semimetal*, Nature Physics 2022 18:10 **18**, 1228 (2022).

[7] D. B. Szombati, S. Nadj-Perge, D. Car, S. R. Plissard, E. P. A. M. Bakkers, and L. P. Kouwenhoven, *Josephson ϕ 0 -Junction in Nanowire Quantum Dots*, Nature Physics 2016 12:6 **12**, 568 (2016).

[8] I. O. Kulik, *Magnitude of the Critical Jopephson Tunnel Current*, J. Exp. Theor. Phys. (U.S.S.R.) **22**, 1211 (1966).

[9] L. N. Bulaevskii, V. V Kuzii, and A. A. Sobyanin, *Superconducting System with Weak Coupling to the Current in the Ground State*, JETP Lett. **25**, 290 (1977).

[10] B. I. Spivak and S. A. Kivelson, *Negative Local Superfluid Densities: The Difference between Dirty Superconductors and Dirty Bose Liquids*, Phys Rev B **43**, 3740 (1991).

[11] R. Avriller and F. Pistolesi, *Andreev Bound-State Dynamics in Quantum-Dot Josephson Junctions: A Washing Out of the 0-π Transition*, Phys Rev Lett **114**, 037003 (2015).

[12] X. Q. Wang, G. Y. Yi, and W. J. Gong, *Effect of Interdot Coulomb Interaction on the Josephson Phase Transition in a Double-Quantum-Dot Junction*, Superlattices Microstruct **109**, 366 (2017).

[13] B. Probst, F. Domínguez, A. Schroer, A. L. Yeyati, and P. Recher, *Signatures of Nonlocal Cooper-Pair Transport and of a Singlet-Triplet Transition in the Critical Current of a Double-Quantum-Dot Josephson Junction*, Phys Rev B **94**, 155445 (2016).

[14] D. Razmadze, R. S. Souto, L. Galletti, A. Maiani, Y. Liu, P. Krogstrup, C. Schrade, A. Gyenis, C. M. Marcus, and S. Vaitiekėnas, *Supercurrent Reversal in Ferromagnetic Hybrid Nanowire Josephson Junctions*, Phys Rev B **107**, L081301 (2023).

[15] J. R. Petta, A. C. Johnson, J. M. Taylor, E. A. Laird, A. Yacoby, M. D. Lukin, C. M. Marcus, M. P. Hanson, and A. C. Gossard, *Coherent Manipulation of Coupled Electron Spins in Semiconductor Quantum Dots*, Science (1979) **309**, 2180 (2005).

[16] F. H. L. Koppens, C. Buizert, K. J. Tielrooij, I. T. Vink, K. C. Nowack, T. Meunier, L. P. Kouwenhoven, and L. M. K. Vandersypen, *Driven Coherent Oscillations of a Single Electron Spin in a Quantum Dot*, Nature **442**, 766 (2006).

[17] C. Barthel, D. J. Reilly, C. M. Marcus, M. P. Hanson, and A. C. Gossard, *Rapid Single-Shot Measurement of a Singlet-Triplet Qubit*, Phys Rev Lett **103**, 160503 (2009).

[18] K. D. Petersson, C. G. Smith, D. Anderson, P. Atkinson, G. A. C. Jones, and D. A. Ritchie, *Charge and Spin State Readout of a Double Quantum Dot Coupled to a Resonator*, Nano Lett **10**, 2789 (2010).

[19] M. Nilsson, I.-J. Chen, S. Lehmann, V. Maulerova, K. A. Dick, and C. Thelander, *Parallel-Coupled Quantum Dots in InAs Nanowires*, Nano Lett **17**, 7847 (2017).

[20] H. Potts, I. –J. Chen, A. Tsintzis, M. Nilsson, S. Lehmann, K. A. Dick, M. Leijnse, and C. Thelander, *Electrical Control of Spins and Giant g -Factors in Ring-like Coupled Quantum Dots*, Nature Communications 2019 10:1 **10**, 1 (2019).

[21] R. Debbarma, H. Potts, C. J. Stenberg, A. Tsintzis, S. Lehmann, K. Dick, M. Leijnse, and C. Thelander, *Effects of Parity and Symmetry on the Aharonov–Bohm Phase of a Quantum Ring*, Nano Lett **22**, 334 (2022).

[22] L. Hofstetter, S. Csonka, J. Nygård, and C. Schönenberger, *Cooper Pair Splitter Realized in a Two-Quantum-Dot Y-Junction*, Nature **461**, 960 (2009).

[23] L. G. Herrmann, F. Portier, P. Roche, A. L. Yeyati, T. Kontos, and C. Strunk, *Carbon Nanotubes as Cooper-Pair Beam Splitters*, Phys Rev Lett **104**, 026801 (2010).

[24] R. S. Deacon, A. Oiwa, J. Sailer, S. Baba, Y. Kanai, K. Shibata, K. Hirakawa, and S. Tarucha, *Cooper Pair Splitting in Parallel Quantum Dot Josephson Junctions*, Nat Commun **6**, 7446 (2015).

[25] A. Bordoloi, V. Zannier, L. Sorba, C. Schönenberger, and A. Baumgartner, *Spin Cross-Correlation Experiments in an Electron Entangler*, Nature **612**, 454 (2022).

[26] J. C. Estrada Saldaña, A. Vekris, G. Steffensen, R. Žitko, P. Krogstrup, J. Paaske, K. Grove-Rasmussen, and J. Nygård, *Supercurrent in a Double Quantum Dot*, Phys Rev Lett **121**, 257701 (2018).

[27] A.Vekris, J. C. Estrada Saldaña, T. Kanne, M. Marnauza, D. Olsteins, F. Fan, X. Li, T. Hvid-Olsen, X. Qiu, H. Xu et al., *Josephson Junctions in Double Nanowires Bridged by In-Situ Deposited Superconductors*, Phys Rev Res **3**, 033240 (2021).

[28] R. Debbarma, M. Aspegren, F. V. Boström, S. Lehmann, K. Dick, and C. Thelander, *Josephson Current via Spin and Orbital States of a Tunable Double Quantum Dot*, Phys Rev B **106**, L180507 (2022).

[29] D. Bouman, R. J. J. van Gulik, G. Steffensen, D. Pataki, P. Boross, P. Krogstrup, J. Nygård, J. Paaske, A. Pályi, and A. Geresdi, *Triplet-Blockaded Josephson Supercurrent in Double Quantum Dots*, Phys Rev B **102**, 220505(R) (2020).

[30] Z. Scherübl, A. Pályi, and S. Csonka, *Transport Signatures of an Andreev Molecule in a Quantum Dot–Superconductor–Quantum Dot Setup*, Beilstein Journal of Nanotechnology **10**, 363 (2019).

[31] S. Droste, S. Andergassen, and J. Splettstoesser, *Josephson Current through Interacting Double Quantum Dots with Spin–Orbit Coupling*, Journal of Physics: Condensed Matter **24**, 415301 (2012).

[32] K. Grove-Rasmussen, G. Steffensen, A. Jellinggaard, M. H. Madsen, R. Žitko, J. Paaske, and J. Nygård, *Yu–Shiba–Rusinov Screening of Spins in Double Quantum Dots*, Nature Communications 2018 9:1 **9**, 1 (2018).

[33] R. Maurand, T. Meng, E. Bonet, S. Florens, L. Marty, and W. Wernsdorfer, *First-Order 0-π Quantum Phase Transition in the Kondo Regime of a Superconducting Carbon-Nanotube Quantum Dot*, Phys Rev X **2**, 011009 (2012).

[34] W. G. van der Wiel, S. De Franceschi, J. M. Elzerman, T. Fujisawa, S. Tarucha, and L. P. Kouwenhoven, *Electron Transport through Double Quantum Dots*, Rev Mod Phys **75**, 1 (2002).

[35] J.-P. Cleuziou, W. Wernsdorfer, V. Bouchiat, T. Ondarçuhu, and M. Monthioux, *Carbon Nanotube Superconducting Quantum Interference Device*, Nat Nanotechnol **1**, 53 (2006).

[36] A. V. Rozhkov and D. P. Arovas, *Josephson Coupling through a Magnetic Impurity*, Phys Rev Lett **82**, 2788 (1999).

[37] E. J. H. Lee, X. Jiang, M. Houzet, R. Aguado, C. M. Lieber, and S. De Franceschi, *Spin-Resolved Andreev Levels and Parity Crossings in Hybrid Superconductor–Semiconductor Nanostructures*, Nat Nanotechnol **9**, 79 (2014).

[38] *See Supplemental Material at Http://Link.Aps.Org/Supplemental/XXXX/XXXXX for Additional Experimental Results and Details on the Modelling*.

[39] H. Potts, M. Leijnse, A. Burke, M. Nilsson, S. Lehmann, K. A. Dick, and C. Thelander, *Selective Tuning of Spin-Orbital Kondo Contributions in Parallel-Coupled Quantum Dots*, Phys Rev B **101**, 115429 (2020).

[40] L. N. Bulaevskii, V. V. Kuzii, A. A. Sobyanin, and P. N. Lebedev, *On Possibility of the Spontaneous Magnetic Flux in a Josephson Junction Containing Magnetic Impurities*, Solid State Commun **25**, 1053 (1978).

[41] H. Sickinger, A. Lipman, M. Weides, R. G. Mints, H. Kohlstedt, D. Koelle, R. Kleiner, and E. Goldobin, *Experimental Evidence of a φ Josephson Junction*, Phys Rev Lett **109**, 107002 (2012).

[42] E. Strambini, A. Iorio, O. Durante, R. Citro, C. SanzFernández, C. Guarcello, I. V. Tokatly, A. Braggio, M. Rocci, N. Ligato et al., *A Josephson Phase Battery*, Nat Nanotechnol **15**, 656 (2020).

[43] D. Z. Haxell, M. Coraiola, M. Hinderling, S. C. ten Kate, D. Sabonis, A. E. Svetogorov, W. Belzig, E. Cheah, F. Krizek, R. Schott et al., *Demonstration of Nonlocal Josephson Effect in Andreev Molecules*, arXiv:2306.00866.